\documentclass[notitlepage,a4,10.5pt]{article}%,twocolumn]{revtex4}

\usepackage{graphicx}

\usepackage{amsmath}%
\usepackage{amsfonts}%
\usepackage{amssymb}%
\usepackage{mathrsfs}
\usepackage{amsthm}
\usepackage{float}

\usepackage{authblk}
\usepackage{enumitem}

\usepackage[left=2.5cm,right=2.5cm,top=3cm,bottom=3cm]{geometry} 

\usepackage{xcolor}
\usepackage{appendix}
\newcommand{\mc}{\mathcal}
\newcommand{\cp}{\times}

\newcommand{\cu}{\nabla\times}

\newcommand{\bol}{\boldsymbol}

\newcommand{\abs}[1]{\left\lvert{#1}\right\rvert}

\newcommand{\lr}[1]{\left({#1}\right)}
\newcommand{\lrs}[1]{\left[{#1}\right]}
\newcommand{\lrc}[1]{\left\{{#1}\right\}}

\newcommand{\p}{\partial}

\newcommand{\ti}[1]{\textit{#1}}
\newcommand{\tb}[1]{\textbf{#1}}

\newcommand{\eq}[1]{\begin{equation}\begin{split}{#1}\end{split}\end{equation}}
\newcommand{\sys}[2]{\begin{subequations}\begin{align}{#1}\end{align}\label{#2}\end{subequations}}

\begin{document}

%\title{Hydrodynamic Approach to the Schr\"odinger-Newton Equation\\ for Spinning Self-Gravitating Fluids}
\title{
Quantum-Fluid Correspondence in Relativistic Fluids with Spin: From Madelung Form to Gravitational Coupling
}
%\title{Quantum-Fluid Correspondence via Madelung Transformation \\for Spinning Self-Gravitating Fluids}
\author[1]{Naoki Sato} %\author[2]{Zhisong Qu} \author[3]{David Pfefferl\'e} \author[2]{Robert L. Dewar} 
\affil[1]{National Institute for Fusion Science, \protect\\ 322-6 Oroshi-cho Toki-city, Gifu 509-5292, Japan \protect\\ Email: sato.naoki@nifs.ac.jp}
%\affil[2]{Mathematical Sciences Institute, The Australian National University, Canberra, ACT 2601, Australia}
%\affil[3]{The University of Western Australia, 35 Stirling Highway, Crawley WA 6009, Australia}
\date{\today}
\setcounter{Maxaffil}{0}
\renewcommand\Affilfont{\itshape\small}

%\twocolumn[
 % \begin{@twocolumnfalse}
    \maketitle
    \begin{abstract}
   % This paper investigates the 
   % the quantum-fluid correspondence for a charged relativistic fluid with intrinsic spin. 
   % We first examine the nonrelativistic case, demonstrating that the inclusion of spin induces a `quantum' correction to the classical fluid energy. This correction, when coupled with Maxwell's equations, naturally gives rise to the Schr\"odinger equation in Madelung form. Building on this, we extend the formalism to a relativistic perfect fluid, identifying the system’s stress-energy-momentum tensor. Our analysis reveals that the trace of the quantum correction to this tensor corresponds to the energy density of an oscillator, with its frequency determined by the vorticity of the spin motion. Then, we utilize the stress-energy-momentum tensor to derive the relationship between the Ricci scalar curvature, as governed by the Einstein field equations, and the fluid density in a static, spherically symmetric configuration. 
   % Lastly, we generalize the Madelung transformation to compressible Navier-Stokes flows with vorticity and viscosity by developiing a tailored Clebsch representation of the velocity field.
   % This theoretical framework offers potential applications for studying fluid-like systems with internal rotational degrees of freedom, which are commonly encountered in astrophysical contexts.  
   This paper explores the quantum-fluid correspondence in a charged relativistic fluid with intrinsic spin. We begin by examining the nonrelativistic case, showing that the inclusion of spin introduces a quantum correction to the classical fluid energy. This correction, coupled with Maxwell's equations, naturally leads to the Schrödinger equation in Madelung form. Building on this foundation, we extend the formalism to a relativistic perfect fluid, identifying the system's stress-energy-momentum tensor. Our analysis reveals that the trace of the quantum correction to this tensor corresponds to the energy density of an oscillator, with its frequency determined by the vorticity of the spin motion. We then use the stress-energy-momentum tensor to establish the relationship between the Ricci scalar curvature, as dictated by the Einstein field equations, and the fluid density in a static, spherically symmetric configuration. Lastly, we generalize the Madelung transformation to compressible Navier-Stokes flows with vorticity and viscosity by developing a tailored Clebsch representation of the velocity field. This theoretical framework offers potential applications for studying fluid-like systems with internal rotational degrees of freedom, commonly encountered in astrophysical settings.
    \end{abstract}
%\vspace{5mm}
%\end{@twocolumnfalse}
%  ]

\section{Introduction}

The Madelung representation \cite{Madelung1, Madelung2} of the wave function is well known for providing a hydrodynamic interpretation of the Schr\"odinger equation for quantum particles. In this paper, we explore the inverse question: can a fluid system with internal rotational (spin) degrees of freedom be described by the Schr\"odinger equation? More specifically, we investigate whether such a fluid can exhibit quantum-like behavior, with its dynamics governed by the Schr\"odinger equation. Our focus is on deriving the governing equation from fundamental fluid mechanical principles, without introducing an assumed `quantum' correction to account for spin dynamics a priori. 

This problem is of practical relevance in astrophysical systems, which can often be viewed as a superposition of fluid parcels, each possessing both electric charge and angular momentum. For instance, a galaxy can be modeled as a charged, spinning fluid, with stars, gas clouds, and planets acting as the constituent fluid elements. 
 Fluid models are also crucial in high-energy heavy-ion collisions and the study of quark-gluon plasmas \cite{Nakamura23, Flor18}. 
Moreover, just as the Madelung representation forms the foundation of the nonlocal, deterministic de Broglie-Bohm interpretation of quantum mechanics \cite{Bohm1, Bohm2}, a quantum-mechanical formulation of such fluid systems could provide valuable theoretical insights into the coupling between gravity and spin degrees of freedom in general fluid dynamics. This framework may have significant implications for understanding the interplay of fundamental forces in astrophysical and cosmological contexts.
Additionally, in much the same way quantum computing has been proposed as a potential method to tackle fundamental problems in Navier-Stokes turbulence \cite{Gaitan, Succi,Meng}, a hydrodynamic theory of fluids with spin could provide a new approach to simulating quantum systems or even an alternative way of implementing quantum computers. The concept of a `fluid computer' \cite{Cardona, Moore}, for example, highlights the potential for fluid systems to be harnessed as computational models.

The hydrodynamic formalism for the Schrödinger equation, originally introduced by Madelung \cite{Madelung1, Madelung2}, can be extended to other relativistic and nonrelativistic quantum equations, such as the Klein-Gordon, Pauli, and Dirac equations. This approach has been thoroughly explored by Takabayasi in a series of works \cite{Takabayasi52, Takabayasi53, Takabayasi54, Takabayasi55, Takabayasi56}. Hydrodynamic variables offer a powerful framework for analyzing quantum systems from a fluid dynamics perspective, often revealing new physical insights. For example, these variables are particularly useful for studying vortex structures in quantum mechanics \cite{Birula, Ovchinnikov, Yoshida}.

In the hydrodynamic formulation, the complex wave function \( \Psi(\bol{x}, t) \) is decomposed into a density \( \rho = |\Psi|^2 \) and a phase \( A \), according to \( \Psi = \sqrt{\rho} \exp(i A / \hbar) \), where \( \hbar \) is the reduced Planck constant. This transformation converts the Schrödinger equation into a system of coupled equations formally equivalent to the continuity and momentum equations of a classical fluid. In this analogy, the fluid has a mass density \( m \rho \), a flow velocity \( \bol{u} = \frac{1}{m} \nabla A \), and a potential energy \( V + V_q \), where \( m \) is the particle mass, \( V \) represents an external potential, and \( V_q = -\frac{\hbar^2}{2m} \frac{\Delta \sqrt{\rho}}{\sqrt{\rho}} \) is the so-called quantum potential.

This quantum-fluid correspondence can also be reversed, allowing one to view fluid systems as quantum systems, which is the focus of the present study.

Using the expression of the Pauli current in the hydrodynamic representation, it has been demonstrated that the quantum potential 
$V_q$  for a massive charged particle with spin can be interpreted as the kinetic energy of an internal velocity associated with spin motion, observed in the center-of-mass frame \cite{Salesi, Recami, Esposito}. This internal dynamics, referred to as zitterbewegung, has been proposed as the underlying mechanism responsible for the electron's spin and magnetic moment \cite{Hestenes}.

In the first part of this paper, we extend the kinetic interpretation of spin in particles to a fluid system with internal rotation, presenting a hydrodynamic derivation of the Schr\"odinger equation. This derivation is formulated for a massive, charged fluid in the absence of external electromagnetic fields, under the assumption that spin represents the angular momentum generated by the rotation of charge within a fluid parcel. The rotational velocity is determined via the fourth Maxwell equation, which relates it to the curl of the magnetic field produced by the spinning charge. The corresponding kinetic energy is identified as the quantum potential and is used to evaluate the total energy and momentum equation of the fluid. 
The Schr\"odinger equation is then derived by applying the inverse of the Madelung transformation. Furthermore, we identify the system's noncanonical Hamiltonian structure, which exhibits the same Poisson bracket formalism as that of the ideal Euler equations in fluid mechanics \cite{Morrison1, Morrison2, Morrison3}.

The derivation of the Schrödinger equation presented in this study for a macroscopic fluid system with spin can also be understood as a fluid-mechanical derivation of the Schrödinger equation for a massive charged particle with spin, once the kinetic interpretation of spin is adopted. In this framework, the Schrödinger equation is not merely an ansatz of quantum theory but rather emerges from fundamental mechanical principles. 

However, it is essential to note that while this construction can be generalized to couple gravity and spin in a macroscopic fluid system, it does not extend to the development of a theory of quantum gravity without conflicting with the standard interpretation of quantum mechanics \cite{Carlip}. These limitations deserve a brief review. 

As in semiclassical quantum gravity, where the classical stress-energy-momentum tensor is replaced by the expectation value \( \langle \Psi | \hat{T}_{\mu\nu} | \Psi \rangle \) of a stress-energy-momentum tensor operator \( \hat{T}_{\mu\nu} \) in a relevant quantum state \( | \Psi \rangle \) on the right-hand side of Einstein's field equations (see, e.g., \cite{Kibble, Wald, Horowitz}), in a hydrodynamic theory of quantum gravity, the wave function transcends its usual probabilistic interpretation and gains physical reality by directly influencing the spacetime metric. 

Moreover, such a theory would be intrinsically nonlinear since the spacetime metric depends on the wave function, while the wave function's evolution, in turn, is affected by the geometry. This nonlinearity contrasts with the linearity of quantum mechanics. These concepts can be illustrated by considering the nonrelativistic limit of such a theory, the Schrödinger-Newton equation \cite{Diosi, Moroz, Tod, Mocz}:
\eq{
{\rm i}\hbar \frac{\partial \Psi}{\partial t} = -\frac{\hbar^2}{2m} \Delta \Psi + m\Phi \Psi, \quad \Delta \Phi = 4 \pi G m |\Psi|^2,\label{SN}
}
where \( \Phi \) denotes the Newtonian gravitational potential. The gravitational nonlinearity in this equation suggests a connection to the spatial localization of the wave function, a feature unattainable in free-particle solutions of the standard Schrödinger equation \cite{Diosi}.

A further challenge for a hydrodynamic theory of quantum gravity arises from the nature of measurement, particularly in relation to the collapse of the wave function. The instantaneous collapse of a quantum state would lead to a discontinuous change in the Einstein tensor, posing difficulties for the consistency of such a theory.

In the second part of the paper, we shift to a relativistic framework and first derive the stress-energy-momentum tensor for the fluid system with spin. We find that the trace of the quantum correction to the classical perfect fluid stress-energy-momentum tensor corresponds to the energy density of an oscillator, with its frequency determined by the vorticity of the internal rotational velocity associated with spin. By substituting this expression into the Einstein field equations, we establish the relationship between the Ricci scalar curvature and the fluid density in a static, spherically symmetric configuration.

For completeness, we note that quantum corrections to the Schwarzschild exterior solution of the Einstein field equations have been explored within the framework of quantum field theory \cite{Duff, Bje}.

In the third part of the paper, we present a generalization of the Madelung transformation to compressible Navier-Stokes flows with non-zero vorticity and viscosity, utilizing a tailored Clebsch representation of the fluid flow. This representation facilitates the construction of multiple wave functions that encapsulate the system’s dynamics.

The paper is structured as follows: In Section 2, we derive the Schrödinger equation from a hydrodynamic model of a perfect fluid with spin. Section 3 focuses on obtaining the stress-energy-momentum tensor for the fluid and interpreting it in terms of the vorticity of spin motion. In Section 4, we apply the derived tensor to the Einstein field equations to establish the relationship between spacetime curvature and fluid density. Section 5 discusses the aforementioned generalization of the Madelung transformation to fluid flows with vorticity and viscosity. Concluding remarks are provided in Section 6.

Throughout the paper, summation over repeated indices is assumed, and we adopt a metric signature of \( (-,+,+,+) \).

\section{Hydrodynamic derivation of the Schr\"odinger equation}
The goal of this section is twofold: first, to derive the Schrödinger equation for a nonrelativistic fluid with spin, consisting of fluid parcels with mass \( m \) and charge \( q \), based on a hydrodynamic model that assumes the spin motion originates from the internal rotation of electric charge; and second, to describe the noncanonical Hamiltonian structure of the system.

\subsection{Derivation of the Schr\"odinger equation}

In the fluid description, the mass density is given by \( m \rho \), where \( \rho(\bol{x}, t) \) represents the number density of fluid parcels at position \( \bol{x} \in \Omega \) and time \( t \in [0, +\infty) \), with \( \Omega \subset \mathbb{R}^3 \) being a smooth, bounded domain. We assume the normalization \( \int_{\Omega} \rho \, dV = 1 \), where \( dV \) is the volume element in \( \mathbb{R}^3 \). Let \( \boldsymbol{v}_s(\boldsymbol{x}, t) \) represent the local spinning velocity of the charged fluid, which describes the internal rotation within a fluid parcel (not to be confused with the fluid velocity \( \boldsymbol{u}(\boldsymbol{x}, t) \)). 
Figure 1 presents a schematic representation of a fluid parcel along with the associated velocity fields.
\begin{figure}[h]
\hspace*{-2.5cm}\centering
\includegraphics[scale=0.35]{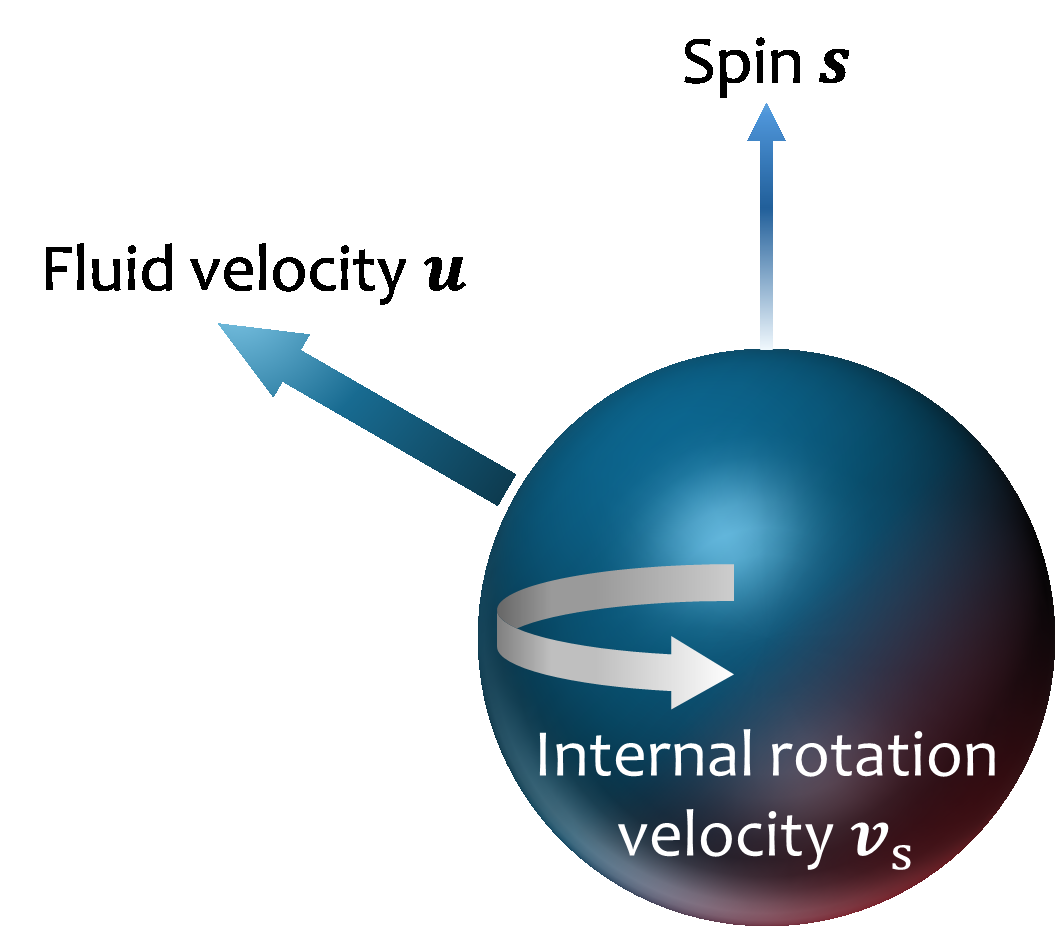}%0.220
\caption{\footnotesize 
Schematic view of a fluid parcel, the fluid velocity $\bol{u}$, the internal velocity $\bol{v}_s$, and the resulting spin $\bol{s}$.}
\label{fig1}
\end{figure}	
The magnetic field generated by the electric current $\bol{J}=q\rho\bol{v}_s$ is determined by the fourth Maxwell equation,
\begin{equation}
\nabla\cp\bol{B}=\mu_0\lr{\bol{J}+\epsilon_0\frac{\p\bol{E}}{\p t}},
\end{equation}
where $\mu_0$ is the vacuum permeability, $\epsilon_0$ the vacuum permittivity, $\bol{B}\lr{\bol{x},t}$ the magnetic field, and $\bol{E}\lr{\bol{x},t}$ the electric field. 
As in magnetohydrodynamics (MHD) \cite{Freid}, we assume that the phase velocities of the relevant electromagnetic waves are significantly slower than the speed of light, and that the characteristic velocities associated with both fluid and spin motion are much smaller than the speed of light. Under these conditions, the displacement current can be neglected, allowing us to set  $\p\bol{E}/\p t=\bol{0}$. This implies
\begin{equation}
\bol{v}_s=\frac{1}{\mu_0 q}\frac{\nabla\cp\bol{B}}{\rho}.\label{vs0}
\end{equation}
In the following, we will denote the spin of a fluid parcel with
\begin{equation}
\bol{S}=\hbar\bol{s},
\end{equation}
where $\bol{s}\lr{\bol{x},t}$ is a vector field with norm $\abs{\bol{s}}=s$ and $\hbar$ is the reduced Planck constant. 
At each spatial position the local spin density is therefore $\hbar\rho\bol{s}$.  
Since the spin density is proportional to the local magnetic moment, it causes a torque
\begin{equation}
\bol{\tau}\propto\hbar\rho\bol{s}\times\bol{B}.
\end{equation}
Let us assume that the relaxation of spin motion within a fluid parcel occurs over a timescale \( \tau_s \), which is much shorter than the characteristic timescale \( \tau_c \) of the fluid system, i.e., \( \tau_s \ll \tau_c \). On the fluid timescale, the spin motion rapidly reaches equilibrium, where the torque \( \boldsymbol{\tau} \) vanishes. Thus,
\eq{
\boldsymbol{B} = \kappa \mu_0 \hbar \rho \boldsymbol{s},
}
for some proportionality coefficient \( \kappa \in \mathbb{R} \), and where the constant \( \mu_0 \) is introduced for convenience. 

Now observe that the local spin density \( \hbar \rho \boldsymbol{s} \), which represents angular momentum density, must be proportional to the particle mass \( m \) but independent of the charge \( q \), while the magnetic field \( \boldsymbol{B} \) must be proportional to the electric charge \( q \) but independent of \( m \). Moreover, the ratio
\eq{
\frac{B}{\mu_0 \hbar \rho s},
}
where \( B = |\boldsymbol{B}| \), has dimensions of charge over mass. Therefore, we expect \( \kappa = |g| q / 2m \), so that
\eq{
\boldsymbol{B} = \frac{|g|}{2} \frac{q \mu_0 \hbar}{m} \rho \boldsymbol{s}, \label{BS}
}
where \( g \) is a dimensionless constant. 

It is useful to note that the experimental value of this constant is approximately \( -2 \) for spin-\( \frac{1}{2} \) particles such as the electron, yielding \( \kappa = q/m \). To simplify the exposition, and without loss of generality, we will absorb the factor \( |g| / 2 \) into the reduced Planck constant, so that \( |g| \hbar / 2 \rightarrow \hbar \).

The velocity \eqref{vs0} thus evaluates as

\begin{equation}
\bol{v}_s=\frac{\hbar}{m}\lr{\nabla\log\rho\times\bol{s}+\nabla\times\bol{s}},\label{vs}
\end{equation}
while the kinetic energy associated with $\bol{v}_s$ is
\begin{equation}
K_{\Omega}^s=\frac{m}{2}\int_{\Omega}\rho\bol{v}_s^2\,dV=\frac{\hbar^2}{2m}\int_{\Omega}\rho\left[\abs{\nabla\log\rho}^2\bol{s}^2-\lr{\nabla\log\rho\cdot\bol{s}}^2+2\nabla\log\rho\times\bol{s}\cdot\nabla\times\bol{s}+\lr{\cu{\bol{s}}}^2\right]\,dV.
\end{equation}
To derive the Schrödinger equation, it is sufficient to consider a regime in which the spin \( \boldsymbol{s} \) is a spatially constant vector and the second term on the right-hand side, \( \nabla \log \rho \cdot \boldsymbol{s} \), is negligible, i.e., \( \nabla \log \rho \cdot \boldsymbol{s} \sim 0 \). This condition, which implies that density gradients tend to be perpendicular to the particle spin, occurs, for example, in a two-dimensional flow where the spin vector points out of the plane. 

It is important to note, however, that this assumption is not strictly necessary for the validity of the theory. If the additional terms are retained, they contribute extra terms to the Schrödinger equation, potentially leading to more complex dynamics.

%is expected within the nonrelativistic limit of the Dirac equation (see e.g. \cite{Salesi,Recami}). 
Maintaining the analogy with spin-\( 1/2 \) particles, we set \( s^2 = 1/4 \). Naturally, different values can be chosen as needed to model the magnitude of the intrinsic spin. Consequently, we obtain
\begin{equation}
K_{\Omega}=\frac{\hbar^2}{8m}\int_{\Omega}\rho\abs{\nabla\log\rho}^2\,dV.\label{K12}
\end{equation}
The potential energy \( V_q \), acting on the fluid due to the kinetic energy of spin motion \( K_{\Omega} \) (as given in equation \eqref{K12}), can be determined by considering the change in \( K_{\Omega} \) caused by a variation \( \delta \rho \) in the density profile. Assuming $\delta\rho$ to vanish on the boundary $\p\Omega$, this gives:
\eq{
\delta K_{\Omega} = \int_{\Omega} V_q \, \delta \rho \, dV,
}
and 
\eq{
V_q = \frac{\delta K_{\Omega}}{\delta \rho} = \frac{\hbar^2}{4m} \left( \frac{1}{2} \left| \nabla \log \rho \right|^2 - \frac{\Delta \rho}{\rho} \right) = -\frac{\hbar^2}{2m} \frac{\Delta \sqrt{\rho}}{\sqrt{\rho}}.\label{QP}
}
Notice that the potential \( V_q \), which describes the effect of intrinsic rotation, is mathematically equivalent to the quantum potential encountered in the Madelung formulation of the standard Schrödinger equation for a nonrelativistic particle. In the following, we will refer to \( V_q \) as the `quantum potential'.

The continuity and momentum equations for the fluid thus take the form:
\begin{subequations}
\begin{align}
    \frac{\partial \rho}{\partial t} &= -\nabla \cdot (\rho \boldsymbol{u}), \\
    \frac{\partial \boldsymbol{u}}{\partial t} &= -\boldsymbol{u} \cdot \nabla \boldsymbol{u} + \nabla \left( \frac{\hbar^2}{2m^2} \frac{\Delta \sqrt{\rho}}{\sqrt{\rho}} - \frac{V}{m} \right) - \frac{1}{m \rho} \nabla P,
\end{align} \label{Seq}
\end{subequations}
where \( \boldsymbol{u}(\boldsymbol{x}, t) \) represents the fluid velocity, \( V(\boldsymbol{x}) \) denotes an external potential energy (a self-induced gravitational potential energy may also be considered, as discussed in Section 3), and \( P(\rho) \) refers to a barotropic pressure field that depends on the fluid density \( \rho \).
As is customary in fluid mechanics, it is useful to introduce a function \( \eta(\rho) \) such that
\eq{
\rho \nabla \eta = \nabla P,
}
so that system \eqref{Seq} takes the form:
\begin{subequations}
\begin{align}
    \frac{\partial \rho}{\partial t} &= -\nabla \cdot (\rho \boldsymbol{u}), \\
    \frac{\partial \boldsymbol{u}}{\partial t} &= -\boldsymbol{u} \cdot \nabla \boldsymbol{u} + \nabla \left( \frac{\hbar^2}{2m^2} \frac{\Delta \sqrt{\rho}}{\sqrt{\rho}} - \frac{V}{m} - \frac{\eta}{m} \right).\label{Seqbx}
\end{align} \label{Seqx}
\end{subequations}
System \eqref{Seqx} describes the Euler equations for a fluid with spin.
We remark that in \eqref{Seqx}, we have omitted electromagnetic forces to simplify the analysis, assuming they are negligible compared to other terms. However, it is straightforward to generalize this theory to a magnetohydrodynamics (MHD) framework with spin.

System \eqref{Seqx} corresponds to the Madelung form of the Schr\"odinger equation, which can be obtained 
in the case of irrotational flows,
\begin{equation}
\bol{u}=\frac{1}{m}\nabla A,\label{uA}
\end{equation}
with $A\lr{\bol{x},t}$ a scalar function. Indeed, after substitution of \eqref{uA} system \eqref{Seqx} becomes
\begin{subequations}
\begin{align}
\frac{\p\rho}{\p t}&=-\nabla\cdot\lr{\rho\frac{\nabla A}{m}},\\
\frac{\p A}{\p t}&=-\frac{\abs{\nabla A}^2}{2m}+\frac{\hbar^2}{2m}\frac{\Delta\sqrt{\rho}}{\sqrt{\rho}}-V-\eta,\label{Seq2b}
\end{align}\label{Seq2}
\end{subequations}
where in \eqref{Seq2b} a spatial constant of integration $\p f\lr{t}/\p t$ has been absorbed in the definition of $A$ according to $A\rightarrow A-f$.  
Defining the complex wave function 
\begin{equation}
\Psi=\sqrt{\rho}\exp\left\{\frac{{\rm i}}{\hbar}A\right\},
\end{equation}
system \eqref{Seq2} thus gives the (nonlinear) Schr\"odinger equation
\begin{equation}
{\rm i}\hbar\frac{\p\Psi}{\p t}=-\frac{\hbar^2}{2m}\Delta\Psi+\lrs{V+\eta\lr{\abs{\Psi}}}\Psi.\label{Sch}
\end{equation}
Observe that equation \eqref{Sch} is nonlinear, 
with the linear cases occurring when $\eta$ is a spatial constant.  
We remark that, however, equation \eqref{Seqx} is more general than \eqref{Sch} since
the derivation carried out above does not constrain the fluid velocity $\bol{u}$ to be curl-free. In section 5 we will see how fluid flows with both vorticity and viscosity
can still be cast in quantum form by 
introducing a Clebsch representation of the velocity field and 
a corresponding set of multiple wave functions.

A few words are in order regarding the physical interpretation of the fluid-quantum correspondence and the model equations \eqref{Seqx}, particularly if one applies the kinetic interpretation of spin to the case of a quantum particle instead of a macroscopic fluid with spin. Specifically, the occurrence of irrotational flows may be a consequence of viscous effects associated with intrinsic rotation. In this interpretation, vortices are progressively dissipated, leading to the onset of potential flows. Thus, the Schrödinger equation \eqref{Sch} could arise from a dissipation mechanism—represented by a diffusion term added to the right-hand side of \eqref{Seqbx}—that erodes nonlinearities over short timescales.

The fluid analogy suggests a diffusion term \( \nu_q \Delta \boldsymbol{u} \) for solenoidal flows where \( \nabla \cdot \boldsymbol{u} = 0 \), with \( \nu_q \) being a quantum kinematic viscosity coefficient. To ensure that the corresponding momentum changes occur more rapidly than those induced by the quantum potential \( V_q \), we require \( \frac{m \nu_q}{\hbar} \gg 1 \). This viscosity can be interpreted as arising from dissipation mechanisms associated with internal rotation.

Additionally, it should be noted that the fluid form \eqref{Seqx} of the Schr\"odinger equation suggests the existence of an underlying kinetic description of the dynamics. Indeed, fluid models emerge as coarse-grained representations of kinetic theory. The potential for such a kinetic (or phase space) formulation of quantum mechanics was originally explored by Wigner \cite{Wigner32}, Moyal \cite{Moyal49}, and Takabayashi \cite{Takabayashi54}.

\subsection{Noncanonical Hamiltonian structure}
Let us now determine the Hamiltonian structure of system \eqref{Seqx}. The Hamiltonian (total energy) of the system is given by
\eq{
H_{\Omega} = \int_{\Omega} \rho \left( \frac{1}{2}m \boldsymbol{u}^2 + V + \frac{h(\rho)}{\rho} + \frac{\hbar^2}{8m} |\nabla \log \rho|^2 \right) dV, \label{H}
}
where the function \( h(\rho) \) satisfies \( \eta = {dh}/{d\rho} \). Notice that the last term corresponds to the kinetic energy of intrinsic rotation, \( K_{\Omega} \), as given in equation \eqref{K12}. Since the only difference between system \eqref{Seqx} and the standard Euler equations is the inclusion of the quantum potential \( V_q \), it is reasonable to expect that this system possesses the same phase space structure as a classical ideal fluid. 

Mathematically, this means that the fluid with spin and the classical ideal fluid share the same Poisson algebra. This is indeed the case, as can be verified by substituting the Hamiltonian \eqref{H}, which includes the effect of spin, into the Poisson bracket \( \left\{,\right\} \) of the ideal Euler equations of fluid mechanics:
\eq{
\frac{\partial \rho}{\partial t} = \left\{\rho, H_{\Omega}\right\}, \quad \frac{\partial \boldsymbol{u}}{\partial t} = \left\{\boldsymbol{u}, H_{\Omega}\right\}, \label{Euler}
}
where, denoting the independent phase space variables as \( \boldsymbol{z} = (m\rho, \boldsymbol{u}) \), the Poisson bracket \( \left\{,\right\} \) is defined as:
\eq{
\left\{ F, G \right\} = \int_{\Omega} \frac{\delta F}{\delta \boldsymbol{z}} \cdot \mathcal{J} \frac{\delta G}{\delta \boldsymbol{z}} \, dV.
}
Here, \( F \) and \( G \) are smooth functionals of the phase space variables \( \boldsymbol{z} \), \( \delta \) denotes functional differentiation, and the Poisson operator \( \mathcal{J} \) is given by:
\begin{equation}
\mathcal{J} = \begin{bmatrix} 0 & -\nabla \cdot \\ -\nabla & -\frac{\nabla \times \boldsymbol{u}}{m \rho} \times \end{bmatrix}.
\end{equation}
By setting variations to zero on the boundary \( \partial \Omega \), the result of evaluating \eqref{Euler} is precisely system \eqref{Seqx}.

Since the Hamiltonian structure preserves phase space volume and prevents entropy growth, system \eqref{Seqx} represents a deterministic model. This feature is inherited by the Schrödinger equation \eqref{Sch}, implying that for a standard quantum particle, the Schrödinger equation describes the deterministic evolution of the probability density \( \rho = |\Psi|^2 \).

Finally, it is worth noting that since system \eqref{Seqx} shares the same Poisson algebra as classical ideal fluids, it also possesses the same Casimir invariants, meaning that helicity remains a topological invariant.

\section{Stress-energy-momentum tensor of a perfect fluid with spin}
The purpose of this section is to obtain the stress-energy-momentum tensor for a perfect fluid with spin. 
For simplicity, we shall consider the case in which no external potential is present, $V=0$. 
%The hydrodynamic formulation \eqref{Seq} provides a straightforward way to couple quantum mechanics with general relativity. Indeed, it is sufficient to determine the contribution of the quantum potential \eqref{QP} to the  stress-energy-momentum tensor $T_{\mu\nu}$. In particular, 
We expect to find   
\begin{equation}
T_{\mu\nu}=C_{\mu\nu}+Q_{\mu\nu},\label{Tmn}
\end{equation}
where $C_{\mu\nu}$ corresponds to classical stress-energy while $Q_{\mu\nu}$ describes quantum corrections. 
The spacetime metric of the fluid will then be given by solution of Einstein's field equations
\begin{equation}
R_{\mu\nu}-\frac{1}{2}Rg_{\mu\nu}%+\Lambda g_{\mu\nu}
=\frac{8\pi G}{c^4}T_{\mu\nu},\label{EFE}
\end{equation}
where %$\Lambda$ is the cosmological constant and 
the source on the right-hand side is given by \eqref{Tmn}. 
%intended to include quantum effects. 
Here, $R_{\mu\nu}$ denotes the Ricci curvature tensor, $R=g^{\mu\nu}R_{\mu\nu}$ the Ricci scalar curvature, 
$g_{\mu\nu}$ the metric tensor, $G$ the Newtonian gravitational constant, and $c$ the speed of light. 
%, and $T_{\mu\nu}$ 
%the stress-energy-momentum tensor.

To evaluate the stress-energy-momentum tensor \eqref{Tmn}, 
first recall that 
for a classical perfect fluid with mass density $m\rho$, pressure $P$,  velocity $u^\mu=dx^\mu/d\tau$ such that $u^{\mu}u_{\mu}=-c^2$, and proper time $\tau$, the stress-energy-momentum tensor has expression
\begin{equation}
C_{\mu\nu}^{\rm}=m\rho u_{\mu}u_{\nu}+P\lr{g_{\mu\nu}+\frac{u_{\mu}u_{\nu}}{c^2}}.\label{SEM}
\end{equation}
The corresponding equations of motion follow from the divergence-free nature of the Einstein tensor, which yields
$\nabla_{\nu}T^{\mu\nu}=0$, where $\nabla_{\nu}$ denotes covariant differentiation with respect to $x^\nu$ (on a vector field $Y=Y^\mu\p_\mu$, we have $\nabla_{\nu}Y^\mu=\p_{\nu}Y^\mu+\Gamma_{\nu j}^\mu Y^j$ where $\p_\mu$ and $\Gamma_{\nu j}^\mu$ are the tangent vector in the $\mu$ direction and the Christoffel symbols respectively, while on a contravariant $2$-tensor $T^{\mu\lambda}$ we have $\nabla_\nu T^{\mu\lambda}=\p_\nu T^{\mu\lambda}+\Gamma_{\nu j}^\mu T^{j\lambda}+\Gamma^{\lambda}_{\nu j}T^{\mu j}$). 
Denoting with $\Pi$ the projection operator onto the space perpendicular to the four-velocity $\mc{U}=u^\mu\p_{\mu}$ with components $\Pi_{\mu}^{\nu}=\delta_{\mu}^{\nu}+u_{\mu}u^{\nu}/c^2$, 
and with ${\rm div}$ and ${\rm grad}$ the divergence and gradient operators with respect to the spacetime metric $g_{\mu\nu}$, 
the equations of motion $\nabla_{\nu}T^{\mu\nu}=0$ can be separated into a component parallel to $\mc{U}$ and a component perpendicular to $\mc{U}$ according to \cite{FrankelGR}
\begin{subequations}
\begin{align}
{\rm div}\lr{m\rho\,\mc{U}}&=-\frac{P}{c^2}{\rm div}\lr{\mc{U}},\label{EoM1}\\
\lr{m\rho+\frac{P}{c^2}}\nabla_{\mc{U}}\mc{U}&=-\Pi\,{\rm grad}\lr{P}.\label{EoM2}
\end{align}\label{EoM}
\end{subequations}
%From \eqref{Seq} one sees that the quantum system is not subject to pressure forces. 
Here, it is useful to observe that in the absence of pressure, the tensor $C_{\mu\nu}$ can be simplified as
\begin{equation}
C_{\mu\nu}=m\rho u_{\mu}u_{\nu}, 
\end{equation}
while equations \eqref{EoM1} and \eqref{EoM2} reduce to the continuity equation and the geodesic equation respectively,  
\begin{subequations}
%\begin{equation}
\begin{align}
{\rm div}\lr{\rho\,\mc{U}}&=0,\\\nabla_{\mc{U}}\mc{U}&=\bol{0}.
\end{align}
%\end{equation}
\end{subequations}
We are now left with the task of determining the quantum contribution $Q_{\mu\nu}$ to the stress-energy-momentum tensor.
This amounts to finding an appropriate expression for $Q_{\mu\nu}$ in terms of the density $\rho$ such that
the equations of motion $\nabla_{\nu}T^{\mu\nu}=0$ reduce to equation \eqref{Seq} in the nonrelativistic limit.
The desired result can be achieved by setting
\begin{equation}
Q_{\mu\nu}=\frac{\hbar^2}{2m}\lr{\frac{\p\sqrt{\rho}}{\p x^{\mu}}\frac{\p\sqrt{\rho}}{\p x^{\nu}}-\sqrt{\rho}\frac{\p^2\sqrt{\rho}}{\p x^{\mu}\p x^{\nu}}}=
\frac{\hbar^2}{4m}\lr{\rho\frac{\p\log\rho}{\p x^{\mu}}\frac{\p\log\rho}{\p x^{\nu}}-\frac{\p^2\rho}{\p x^{\mu}\p x^{\nu}}}=-\frac{\hbar^2}{4m}\rho\frac{\p^2\log\rho}{\p x^{\mu}\p x^{\nu}}.\label{Qmn}
\end{equation}
This can be seen by observing that in the special relativistic limit of the Minkowski metric %$ds^2=-c^2dt^2+dx^2+dy^2+dz^2$
$\eta_{\mu\nu}$, we have
\begin{equation}
\nabla_{\nu}Q^{\mu\nu}=\eta^{\mu\alpha}\eta^{\nu\beta}\p_{\nu}Q_{\alpha\beta}=-\frac{\hbar^2}{2m}\eta^{\mu\alpha}\eta^{\nu\beta}\rho\frac{\p}{\p x^{\alpha}}\lr{\frac{1}{\sqrt{\rho}}\frac{\p^2\sqrt{\rho}}{\p x^{\nu}\p x^{\beta}}}=-\frac{\hbar^2}{2m}\eta^{\mu\alpha}\rho\frac{\p}{\p x^{\alpha}}\lr{\frac{\Delta\sqrt{\rho}}{\sqrt{\rho}}-\frac{1}{c^2\sqrt{\rho}}\frac{\p^2\sqrt{\rho}}{\p t^2}}.
\end{equation}
Hence, in the nonrelativistic limit $c\gg 1$, one obtains
\begin{equation}
\lr{\nabla_{\nu}Q^{\mu\nu}}\p_{\mu}=-\frac{\hbar^2}{2m}\rho\nabla\lr{\frac{\Delta\sqrt{\rho}}{\sqrt{\rho}}}=\rho\nabla V_q,\label{divq}
\end{equation}
which is density times the gradient of the quantum potential \eqref{QP}. Using this result, one can verify that the nonrelativistic limit of $\nabla_{\nu}T^{\mu\nu}=0$ is exactly system \eqref{Seq}. 
We remark that expressions analogous to \eqref{Qmn} for the stress-energy-momentum tensor have been discussed by Takabayasi
and others (see e.g. \cite{Takabayasi52, Takabayasi53, Biro}).

In the nonrelativistic limit the trace $\mc{Q}$ of the tensor $Q_{\mu\nu}$ 
can be rearranged in a number of different ways, 
\begin{equation}
\mc{Q}=Q_{\mu\nu}g^{\mu\nu}=\frac{\hbar^2}{4m}\lr{\rho\abs{\nabla\log\rho}^2-\Delta\rho}=
-\frac{\hbar^2}{4m}\rho\Delta\log\rho
=\rho V_q+k=2k-\frac{\hbar^2}{4m}\Delta\rho,
\end{equation}
where $V_q$ is the quantum potential \eqref{QP} and $k=\hbar^2\rho\abs{\nabla\log\rho}^2/{8m}$ the energy density
associated with the kinetic energy of internal rotation \eqref{K12}. 
The physical interpretation of the trace $\mc{Q}$ can be obtained by recalling the expression of the internal charge rotation velocity \eqref{vs}. For a fluid with spin  such that $\nabla\log\rho\cdot\bol{s}\sim 0$, it implies
\begin{equation}
\frac{\bol{s}\times\bol{v}_s}{s^2}=\frac{\hbar}{m}\nabla\log\rho.
\end{equation}
Taking the divergence of both sides, one obtains
\begin{equation}
\mc{Q}=\frac{1}{2}\hbar\omega_{s}\rho,\label{Q2}
\end{equation}
where $\omega_s=2\bol{s}\cdot\nabla\times\bol{v}_s$ is the projection of the vorticity $\nabla\times\bol{v}_s$ 
in the spin direction. Since $\omega_s$ has dimensions of a frequency, 
the trace $\mc{Q}$ is nothing but the energy density of an oscillator with frequency $\omega_s$.

In conclusion, returning to the relativistic case, the total stress-energy-momentum tensor has expression
\begin{equation}
T_{\mu\nu}=m\rho u_{\mu}u_{\nu}+P\lr{g_{\mu\nu}+\frac{u_{\mu}u_{\nu}}{c^2}}+\frac{\hbar^2}{4m}\lr{\frac{\rho_{\mu}\rho_{\nu}}{\rho}-\rho_{\mu\nu}}
,\label{TmnX}
\end{equation}
where in this notation a lower index applied to a scalar function indicates partial differentiation, e.g. $\rho_{\mu}=\p\rho/\p x^{\mu}$. From equation \eqref{Q2} one sees that the trace of \eqref{TmnX} is
\begin{equation}
T=T_{\mu\nu}g^{\mu\nu}=3P-m\rho c^2+\frac{1}{2}\hbar\omega_s\rho,
\end{equation}
with $\omega_s=-\frac{\hbar}{2m}g^{\mu\nu}\frac{\p^2\log\rho}{\p x^{\mu}\p x^{\nu}}$. 
This expression should be compared with the trace of the stress-energy-momentum tensor of a perfect fluid, $3P-m\rho c^2$.
Using \eqref{TmnX}, the general relativistic equations of motion $\nabla_{\nu}T^{\mu\nu}=0$ for a perfect fluid with spin can be written as
\begin{equation}
%\begin{split}
\nabla_{\nu}T^{\mu\nu}=\nabla_{\nu}\lrs{\lr{m\rho +\frac{P}{c^2}}u^{\nu}}u^{\mu}+\lr{m\rho+\frac{P}{c^2}} u^{\nu}\nabla_{\nu}u^{\mu}+\nabla_{\nu}\lr{Pg^{\mu\nu}+Q^{\mu\nu}}=0,\label{EqX1}
%\end{split}
\end{equation}
or equivalently separated into parallel and perpendicular components with respect to $\mc{U}$, 
\begin{subequations}
\begin{align}
{\rm div}{\lr{m\rho\,\mc{U}}}&=-\frac{P}{c^2}{\rm div}\lr{\mc{U}}+\frac{1}{c^2}\langle {\rm Div}\,{{Q}},\mc{U}\rangle,\\
\lr{m\rho+\frac{P}{c^2}}\nabla_{\mc{U}}\mc{U}&=-\Pi~\lr{{\rm grad}P+{\rm Div}\,{Q}},\label{EqX22}
\end{align}\label{EqX2}
\end{subequations}
where ${\rm Div}\,\lr{Pg+Q}=\lrs{\nabla_{\nu}\lr{Pg^{\mu\nu}+Q^{\mu\nu}}}\p_{\mu}={\rm grad} P+{\rm Div}Q$ denotes the tensorial divergence of the tensor $Pg^{\mu\nu}+Q^{\mu\nu}$, and $\langle,\rangle$ 
the scalar product with respect to the metric tensor $g_{\mu\nu}$. Notice that in going from \eqref{EqX1} to \eqref{EqX2} we used the fact that $u_{\mu}u^{\mu}=-c^2$.

The momentum equation in \eqref{EqX22} can be identified with the general relativistic form of the Schr\"odinger-Newton equation \eqref{SN}. To see this, consider the simplified static case $\p_0=0$ in which the four-velocity does not have a spatial component, $\mc{U}=u^0\lr{x^1,x^2,x^3}\p_0$, and set $P=0$. 
%, and the 
%gravitational field is weak, $\Phi/mc^2<<1$. 
Then, equation \eqref{EqX22} can be written as
\begin{equation}
m\rho \lr{u^0}^2\Gamma_{00}^k\p_k=-\Pi\,{\rm Div}\,Q,\label{SN2}
\end{equation}
where $\lr{u^0}^2=-c^2/g_{00}$ and $\Gamma_{00}^k=-\frac{1}{2}g^{km}\frac{\p g_{00}}{\p x^{m}}$. Recalling that $g_{00}$ is related to the classical Newtonian potential energy $m\Phi$ according to $\sqrt{-g_{00}}=c\lr{1+\Phi/c^2}$, taking the nonrelativistic limit $c\gg 1$, and using equation \eqref{divq} to evaluate the right-hand side, equation \eqref{SN2} becomes
\begin{equation}
\nabla\lr{m\Phi+V_q}=\bol{0},\label{42}
\end{equation}
This equation expresses force balance between 
the gravitational force and the quantum force, as occurs in the static case of the Schr\"odinger-Newton equation \eqref{SN}.  This fact can be seen explicitly by  setting $A=0$ and $\Psi=\Psi^*=\sqrt{\rho}$ in \eqref{SN} when  $\p_t=0$.  

It is also useful to observe that, in the special relativistic limit $g_{\mu\nu}=\eta_{\mu\nu}$,
with $\eta_{\mu\nu}$ the Minkowski metric, and in the absence of pressure, $P=0$, solutions of the Klein-Gordon equation 
\begin{equation}
\Box\Psi=\frac{m^2c^2}{\hbar^2}\Psi,\label{KG}
\end{equation}
where $\Box=\eta^{\mu\nu}\p_{\mu}\p_{\nu}$ is the d'Alembert operator, are solutions of equation \eqref{EqX2}.  
To see this, perform the change of variables $\Psi=\sqrt{\rho}\exp\lr{{\rm i}A/\hbar}$, which transforms equation \eqref{KG} into the system
\begin{subequations}
\begin{align}
A^{\nu}A_{\nu}+m^2c^2=&\hbar^2\frac{\Box\sqrt{\rho}}{\sqrt{\rho}},\\
\p_{\nu}\lr{\rho A^{\nu}}=&0,
\end{align}\label{KG2}
\end{subequations}
with $A^{\nu}=\eta^{\mu\nu}A_{\mu}$. 
On the other hand, consider the case in which 
the velocity field is a potential flow, i.e. 
\begin{equation}
\mc{U}=\frac{1}{m}A_{\mu}\eta^{\mu\nu}\p_{\nu}.
\end{equation}
%Here, $A$ is a scalar function. 
Then, 
\begin{equation}
\nabla_{\nu}T^{\mu\nu}=\frac{1}{m}\p_{\nu}\lr{\rho A^{\nu}}A^{\mu}+\frac{1}{2m}\rho\p^{\mu}\lr{A^{\nu}A_{\nu}-\hbar^2\frac{\Box\sqrt{\rho}}{\sqrt{\rho}}},\label{tmunu}
\end{equation}
where we used the identity $A^{\nu}\p_{\nu}A^{\mu}-\frac{1}{2}\p^{\mu}\lr{A^{\nu}A_{\nu}}=0$. 
The quantity \eqref{tmunu} vanishes identically upon substitution of \eqref{KG2}. Here, the notation $\p^{\mu}=g^{\mu\nu}\p_{\nu}$ for the cotangent vector in the $\mu$ direction has been used. 
It is worth observing that if we demand that $\mc{U}$ 
represents a physical flow such that $u_{\mu}u^{\mu}=m^{-2}A_{\mu}A^{\mu}=-c^2$, 
system \eqref{KG2} reduces to a wave equation for $\sqrt{\rho}$ coupled with the continuity equation,
\begin{subequations}
\begin{align}
\Box\sqrt{\rho}=&0,\\
{\rm div}\lr{\rho\,\mc{U}}=&\bol{0}.
\end{align}
\end{subequations}
Finally, note that special relativistic solutions of \eqref{EqX2} are not solutions of the Klein-Gordon equation because in general 
the quantum term ${\rm Div}\, Q=-\frac{\hbar^2}{2m}\rho\,{\rm grad}\lr{\frac{\Box\sqrt{\rho}}{\sqrt{\rho}}}$ has a non-vanishing component along $\mc{U}$ that prevents the separation of the equations of motion
in the form \eqref{KG2}.

\section{Spacetime curvature}
The goal of this section is to derive the equations governing the spacetime metric produced by a perfect fluid with intrinsic spin through the stress-energy-momentum tensor \eqref{TmnX}. Specifically, we aim to express the Ricci scalar curvature $R$ as a function of the density $\rho$. In physical systems, such curvature could be experimentally measured and used to estimate the magnitude of intrinsic rotational effects.

We focus on spherically symmetric and static solutions of the Einstein field equations \eqref{EFE}. For simplicity, we restrict our analysis to a pressureless fluid, i.e., $P = 0$. Using spherical coordinates $\lr{r,\theta,\varphi}$, we assume that $\rho = \rho\lr{r}$ and the four-velocity components $u^{\mu}$ depend only on $r$. The metric is then given by:
\begin{equation}
ds^2 = g_{tt}\lr{r}dt^2 + g_{rr}\lr{r}dr^2 + r^2d\Omega^2, \quad d\Omega^2 = d\theta^2 + \sin^2\theta\, d\varphi^2. \label{ds2}
\end{equation}
Since the system is static, the four-velocity $\mc{U}$ must take the form:
\begin{equation}
\mc{U} = u^{t}\lr{r} \p_t. \label{u1}
\end{equation}
Here, $u^t\lr{r}$ represents the time component of the four-velocity along the tangent direction $\p_t$.
%, where $t = x^0$. 
This configuration is analogous to a Schrödinger field $\Psi = \sqrt{\rho}\exp\lr{{\rm i}A/\hbar}$, with a spatially constant phase $A = A(t)$, since in a nonrelativistic context, the spatial velocity satisfies $\bol{u}=\mc{U} - u^t \p_t = m^{-1} \nabla A$. 

It is worth noting that the static configuration \eqref{u1} is achievable because, unlike the Schrödinger equation, a static equilibrium can be maintained between the quantum potential and the gravitational potential, as in the Schrödinger-Newton equation (recall equation \eqref{42}).

Observing that $\abs{u}^2 = \lr{u^t}^2 g_{tt} = -c^2$ for a physical flow, we obtain:
\begin{equation}
\mc{U} = \frac{c}{\sqrt{-g_{tt}}} \p_t.
\end{equation}
Additionally, the density $\rho$ is defined with respect to the spatial volume element:
\begin{equation}
\sqrt{g_{rr}} r^2 \sin\theta \, dr \, d\theta \, d\varphi.
\end{equation}
Next, we substitute the stress-energy-momentum tensor \eqref{TmnX} into Einstein's field equations \eqref{EFE}. For convenience, we introduce the constants:
\begin{equation}
\kappa = \frac{8\pi G}{c^4}, \quad \epsilon = \frac{\hbar^2}{4m}.
\end{equation}

Then, Einstein's field equations can be written as
\begin{subequations}
\begin{align}
R_{rr}-\frac{1}{2}Rg_{rr}
%+\Lambda g_{rr}
&=-\kappa\epsilon\rho\frac{\p^2\log\rho}{\p r^2}
%\lr{\frac{\rho_{r}^2}{\rho}+\rho_{rr}}
,\\
R_{r\theta}&=0,\\
R_{r\varphi}&=0,\\
R_{\theta\theta}-\frac{1}{2}Rr^2
%+\Lambda r^2
&=0,\\
R_{\theta\varphi}&=0,\\
R_{\varphi\varphi}-\frac{1}{2}Rr^2\sin^2\theta
%+\Lambda r^2 \sin^2\theta
&=0,\\
R_{tr}&=0,\\
R_{t\theta}&=0,\\
R_{t\varphi}&=0,\\
R_{tt}-\frac{1}{2}Rg_{tt}
%+\Lambda g_{tt}
&=-\kappa m c^2\rho g_{tt}.
\end{align}\label{EFE1}
\end{subequations}
Next, recall that the Ricci curvature tensor $R_{ij}$
is related to the Christoffel symbols of the second kind $\Gamma_{ij}^{k}$ according to
\begin{equation}
R_{ij}=\frac{\p\Gamma_{ij}^{k}}{\p x^{k}}-\frac{\p\Gamma_{kj}^{k}}{\p x^{i}}+\Gamma_{k\ell}^{k}\Gamma^{\ell}_{ij}-\Gamma_{i\ell}^{k}\Gamma_{jk}^{\ell},~~~~\Gamma_{ij}^k=\frac{1}{2}g^{km}\lr{\frac{\p g_{mi}}{\p x^j}+\frac{\p g_{mj}}{\p x^i}-\frac{\p g_{ij}}{\p x^m}}.
\end{equation}
For the spacetime metric \eqref{ds2}, the non-vanishing Christoffel symbols are
\begin{equation}
\begin{split}
&\Gamma_{rr}^{r}=\frac{1}{2}\frac{\p\log g_{rr}}{\p r},~~~~\Gamma_{tt}^r=-\frac{1}{2g_{rr}}\frac{\p g_{tt}}{\p r},~~~~\Gamma_{\theta\theta}^{r}=-\frac{r}{g_{rr}},~~~~\Gamma_{\varphi\varphi}^{r}=-\frac{r\sin^2\theta}{g_{rr}},\\
&\Gamma_{tr}^{t}=\frac{1}{2}\frac{\p\log \abs{g_{tt}}}{\p r},~~~~\Gamma_{r\theta}^\theta=\frac{1}{r},~~~~\Gamma_{\varphi\varphi}^{\theta}=-\sin\theta\cos\theta,~~~~\Gamma_{\theta\varphi}^{\varphi}=\frac{\cos\theta}{\sin\theta},~~~~\Gamma_{\varphi r}^\varphi=\frac{1}{r}.
\end{split}
\end{equation}
The components of the Ricci curvature tensor therefore take the form
\begin{subequations}
\begin{align}
R_{rr}&=-\frac{1}{2}\frac{\p^2\log\abs{g_{tt}}}{\p r^2}+\frac{1}{r}\frac{\p\log g_{rr}}{\p r}+\frac{1}{4}\frac{\p\log\abs{g_{tt}}}{\p r}\left(\frac{\p\log g_{rr}}{\p r}-\frac{\p\log\abs{g_{tt}}}{\p r}\right),\\
R_{r\theta}&=0,\\
R_{r\varphi}&=0,\\
R_{\theta\theta}&=1-\frac{\p}{\p r}\lr{\frac{r}{g_{rr}}}-\frac{r}{2g_{rr}}\frac{\p\log\lr{g_{rr}\abs{g_{tt}}}}{\p r},\\
R_{\theta\varphi}&=0,\\
R_{\varphi\varphi}&=\sin^2\theta\, R_{\theta\theta},\\
%-\sin^2\theta\left[
%\frac{\p}{\p r}\lr{\frac{r}{g_{rr}}}+\frac{r}{2g_{rr}}\frac{\p\log\lr{g_{rr}\abs{g_{tt}}}}{\p r}
%\right],\\
R_{tr}&=0,\\
R_{t\theta}&=0,\\
R_{t\varphi}&=0,\\
R_{tt}&=-\frac{1}{2g_{rr}}\frac{\p^2g_{tt}}{\p r^2}-\frac{1}{rg_{rr}}\frac{\p g_{tt}}{\p r}+\frac{1}{4g_{rr}^2}\frac{\p g_{tt}}{\p r}\frac{\p g_{rr}}{\p r}+\frac{1}{4g_{rr}g_{tt}}\lr{\frac{\p g_{tt}}{\p r}}^2.
%-\frac{1}{2}\frac{\p}{\p r}\lr{\frac{1}{g_{rr}}\frac{\p g_{tt}}{\p r}}-\frac{1}{rg_{rr}}\frac{\p g_{tt}}{\p r}-\frac{1}{4g_{rr}^2}\frac{\p g_{tt}}{\p r}\frac{\p g_{rr}}{\p r}+\frac{1}{4g_{rr}g_{tt}}\lr{\frac{\p g_{tt}}{\p r}}^2.
\end{align}\label{Rg}
\end{subequations}
Using these expressions the Ricci scalar curvature can be evaluated as
\begin{equation}
R=\frac{R_{rr}}{g_{rr}}+\frac{R_{tt}}{g_{tt}}+\frac{2R_{\theta\theta}}{r^2}.\label{RicciR}
\end{equation}
%Recalling that $u^{\mu}u_{\mu}=-c^2$, from the second equation in \eqref{Pu} one also obtains
%\begin{equation}
%u_t^2=-g_{tt}\lr{c^2+\frac{\zeta_r^2}{g_{rr}}}.\label{ut2}
%\end{equation}
Hence, using equations \eqref{Rg} and \eqref{RicciR}, %and \eqref{ut2}, 
Einstein's field equations \eqref{EFE1} reduce to three equations that must be solved for the variables $g_{rr}$, $g_{tt}$, and $\rho$:
%, and $\zeta$:
\begin{subequations}
\begin{align}
\frac{1}{2}\lr{\frac{R_{rr}}{g_{rr}}-\frac{R_{tt}}{g_{tt}}}-\frac{R_{\theta\theta}}{r^2}
%+\Lambda
&=-\frac{\kappa\epsilon}{g_{rr}}\
\rho\frac{\p^2\log\rho}{\p r^2}
%rho_r\frac{\p}{\p r}\log\lr{\frac{\p\log\rho}{\p r}}
,\\
\frac{R_{rr}}{g_{rr}}+\frac{R_{tt}}{g_{tt}}
%+\Lambda
&=0,\\
\frac{1}{2}\lr{\frac{R_{rr}}{g_{rr}}-\frac{R_{tt}}{g_{tt}}}+\frac{R_{\theta\theta}}{r^2}
%+\Lambda
&=\kappa mc^2\rho.%,\\
%\frac{\p p}{\p r}&=-\rho\frac{\p}{\p r}\lr{\frac{\Delta_g\sqrt{\rho}}{\sqrt{\rho}}}.
\end{align}\label{R3}
\end{subequations}
%Recall that here the relationship between $\rho$ and $p$ is given by the first equation in \eqref{Pu}, 
%which can also be written as
%\begin{equation}
%\frac{\p p}{\p r}=-\rho\frac{\p}{\p r}\lr{\frac{\Delta_g\sqrt{\rho}}{\sqrt{\rho}}},~~~~\Delta_{g}=\frac{1}{r^2\sqrt{g_{rr}}}\frac{\p}{\p r}\lr{\frac{r^2}{\sqrt{g_{rr}}}\frac{\p}{\p r}}.\label{prho}
%\end{equation}
With the help of the second equation, this system can be simplified to
\begin{subequations}
\begin{align}
\frac{R_{rr}}{g_{rr}}-\frac{R_{\theta\theta}}{r^2}
%+\frac{3}{2}\Lambda
&=-\frac{\kappa\epsilon}{g_{rr}}
\rho\frac{\p^2\log\rho}{\p r^2}
%\rho_r\frac{\p}{\p r}\log\lr{\frac{\p\log\rho}{\p r}}
,\label{chichi}\\
\frac{R_{rr}}{g_{rr}}+\frac{R_{tt}}{g_{tt}}
%+\Lambda
&=0,\label{etaeta}\\
\frac{R_{rr}}{g_{rr}}+\frac{R_{\theta\theta}}{r^2}
%+\frac{3}{2}\Lambda
&=\kappa m c^2\rho.\label{rho}
\end{align}\label{R4}
\end{subequations}
System \eqref{R3} can be further reduced to two independent equations by eliminating $\rho$ with the aid of the third equation, and by substituting the corresponding expression in the first equation. % and the relationship between $p$ and $\rho$ (equation \eqref{prho}). 
%Observing that static solutions must satisfy $\zeta_r=0$, 
The resulting equations for 
the metric coefficients $g_{rr}$ and $g_{tt}$ can be explicitly written as
\begin{subequations}
\begin{align}
\frac{R_{rr}}{g_{rr}}-\frac{R_{\theta\theta}}{r^2}
%+\frac{3}{2}\Lambda
&=-\frac{\epsilon}{mc^2 g_{rr}}
\lr{\frac{R_{rr}}{g_{rr}}+\frac{R_{\theta\theta}}{r^2}}
\frac{\p^2}{\p r^2}\log\lr{\frac{R_{rr}}{g_{rr}}+\frac{R_{\theta\theta}}{r^2}}
%\frac{\p}{\p r}\lr{\frac{R_{rr}}{g_{rr}}+\frac{R_{\theta\theta}}{r^2}
%+\frac{3}{2}\Lambda
%}\frac{\p}{\p r}\log\left[\frac{\p}{\p r}\log\lr{\frac{R_{rr}}{g_{rr}}+\frac{R_{\theta\theta}}{r^2}
%+\frac{3}{2}\Lambda
%}\right]
,\label{R51}\\
\frac{R_{rr}}{g_{rr}}+\frac{R_{tt}}{g_{tt}}
%+\Lambda
&=0.\label{R52}
\end{align}\label{R5}
\end{subequations}
Next, observe that system \eqref{R4} can be used to express the Ricci scalar curvature \eqref{RicciR} in terms of the density $\rho$. Indeed, substituting \eqref{etaeta} into \eqref{RicciR}, one gets \eq{
R=\frac{2R_{\theta\theta}}{r^2}.\label{61}}
Then, using 
equations 
\eqref{61} and   \eqref{rho} to express $R_{rr}$ and $R_{\theta\theta}$ in terms of the Ricci scalar curvature $R$ and the density $\rho$, equation   \eqref{chichi} gives
\begin{equation}
R=%2\frac{R_{\theta\theta}}{r^2}=
\kappa\rho\lr{ mc^2+\frac{\epsilon}{g_{rr}}\frac{\p^2\log\rho}{\p r^2}}.
\end{equation}
Note that this expression relates the spacetime curvature $R$ to the fluid density $\rho$ and  the reduced Planck constant $\hbar$ through the parameter $\epsilon$. 
Next, suppose that the density $\rho$ is approximated by a Gaussian profile $\rho=\rho_0\exp\left\{-r^2/2\sigma^2\right\}$, where $\rho_0$ is a positive real constant and $\sigma$ a positive length parameter characterizing the spatial dimension of the fluid. Then,
\begin{equation}
R=\kappa \rho\lr{mc^2-\frac{\epsilon}{ \sigma^2}g^{rr}}.
\end{equation}
Considering radial distances such that $\sigma/r\ll 1$, we may expand the metric coefficient $g^{rr}$ in powers of the ratio $\sigma/r$ according to $g^{rr}=1+\lambda\sigma/r+O\lr{\lr{\sigma/r}^2}$ with $\lambda\in\mathbb{R}$ so that $g^{rr}\rightarrow 1$ in the limit $r\rightarrow\infty$ where the fluid density $\rho$ vanishes. At leading order for the ratio $R/\rho$, one thus obtains
\begin{equation}
R=\kappa \mc{E}\rho,~~~~\mc{E}=\lr{mc^2-\frac{\hbar^2}{4m\sigma^2}}.
\end{equation}
Hence, at large enough distances the curvature $R$ is roughly proportional to the density $\rho$, with proportionality coefficient $\kappa\mc{E}$ given by the Einstein gravitational constant $\kappa$ times the difference $\mc{E}$ between the rest energy of a fluid parcel $mc^2$ and the characteristic energy of internal rotation $\hbar^2/4m\sigma^2$. 

\section{Viscosity and Vorticity}

The aim of this section is to generalize the Madelung transformation to include flows with vorticity and viscosity. 
Specifically, we want to map the compressible Navier-Stokes equations with spin 
\begin{subequations}
\begin{align}
    & \frac{\partial \rho}{\partial t} = -\nabla \cdot (\rho \boldsymbol{u}), \\
    & \frac{\partial \boldsymbol{u}}{\partial t} = -\boldsymbol{u} \cdot \nabla \boldsymbol{u} + \nabla \left( \frac{\hbar^2}{2m^2} \frac{\Delta \sqrt{\rho}}{\sqrt{\rho}} - \frac{V}{m} - \frac{\eta}{m} \right) + \nu \Delta \boldsymbol{u} + \left( \xi + \frac{\nu}{3} \right) \nabla (\nabla \cdot \boldsymbol{u}),
\end{align}
\label{NS}
\end{subequations}
into an equivalent set of equations for an appropriate set of wave functions. 
Here, $\nu$ and $\xi$ denote the shear kinematic viscosity and the bulk kinematic viscosity respectively, which are assumed to be positive constants.
We first address irrotational viscous flows and then extend the analysis to flows with both vorticity and viscosity.

\subsection{Madelung transform for irrotational viscous  flows}
For irrotational flows \eqref{uA}, system \eqref{NS} can be written as
\sys{
\frac{\p\rho}{\p t}&=-\nabla\cdot\lr{\rho\frac{\nabla A}{m}},\\
\frac{\p A}{\p t}&=-\frac{\abs{\nabla A}^2}{2m}+\frac{\hbar^2}{2m}\frac{\Delta\sqrt{\rho}}{\sqrt{\rho}}-V-\eta+\lr{\frac{4}{3}\nu+\xi}\Delta A,
}{Seqv}
Noting that
\eq{
\Delta A=\frac{\hbar}{2{\rm i}}\nabla\cdot\lr{\frac{\Psi^*\nabla\Psi-\Psi\nabla\Psi^*}{\abs{\Psi}^2}
},
}
where, as usual, $\Psi=\sqrt{\rho}\exp\lrc{{\rm i}A/m}$ is the complex wave function,  
it follows that the `quantum' form of system \eqref{Seqv} is 
\eq{
{\rm i}\hbar\frac{\p\Psi}{\p t}=-\frac{\hbar^2}{2m}\Delta\Psi+\lrs{V+\eta\lr{\abs{\Psi}}}\Psi+\frac{{\rm i}\hbar}{2}\lr{\frac{4}{3}\nu+\xi}\nabla\cdot\lr{\frac{\Psi^*\nabla\Psi-\Psi\nabla\Psi^*}{\abs{\Psi}^2}}.
}

\subsection{Madelung transform for viscous flows with vorticity}
It is well known that the Madelung transformation can be extended to ideal flows (without dissipation) with non-vanishing vorticity by employing a tailored Clebsch representation \cite{YosClebsch, YosMor} of the fluid velocity and a corresponding Pauli-Schrödinger spinor field \cite{Yoshida}. However, the application of the Madelung transformation to flows with dissipation, particularly for fluid flows governed by the Navier-Stokes equations, remains an open problem. 

In this work, we utilize the Clebsch representation developed in \cite{Sato21} for fluid flows with vorticity to construct a Madelung transformation for the Navier-Stokes equations in two dimensions, or the Navier-Stokes equations in three-dimensions with flows such that vorticity locally defines the normal of a two-dimensional surface. Furthermore,
as discussed in more detail at the end of this section, 
we note that this transformation can be extended to fully three-dimensional Navier-Stokes turbulence, albeit with more involved algebra, by further increasing the number of Clebsch potentials 
used to represent
the velocity field. 
%to define an appropriate set of complex wave functions.

Let \( \boldsymbol{u}(\boldsymbol{x}, t) \) denote a velocity field in three spatial dimensions with vorticity \( \boldsymbol{\omega} = \nabla \times \boldsymbol{u} \). We shall develop a Madelung transformation for compressible Navier-Stokes flows satisfying the following two conditions:
\begin{subequations}
\begin{align}
    &\boldsymbol{u} = \frac{1}{m}\nabla A + \frac{p}{m\rho} \nabla q, \label{epi2d} \\
    &\boldsymbol{\omega} \cdot \nabla \times \boldsymbol{\omega} = 0, \label{intvor}
\end{align} \label{2dNS}
\end{subequations}
where \( A(\boldsymbol{x}, t) \), \( p(\boldsymbol{x}, t) \), $\rho\lr{\bol{x},t}$, and \( q(\boldsymbol{x}, t) \) are the Clebsch potentials. Note that \( A \) and \( q \) can be multivalued functions as long as \( \nabla A \) and \( \nabla q \) remain single-valued. The reason for placing the density \( \rho \) below the potential \( p \) is that this choice simplifies the identification of the Madelung transformation (see below).

Three-dimensional flows admitting the Clebsch representation \eqref{epi2d} are called epi-2D flows \cite{YosMor} and are characterized by the property that their total helicity \( \mathcal{H} \) in any smooth bounded domain \( \Omega \subset \mathbb{R}^3 \) with boundary \( \partial \Omega \) and unit outward normal \( \boldsymbol{n} \) on \( \partial \Omega \) is the volume integral of an exact form:
\eq{
\mathcal{H} = \int_{\Omega} \boldsymbol{u} \cdot \nabla \times \boldsymbol{u} \, dV = \int_{\partial \Omega} \frac{p}{m^2\rho} \nabla q \times \nabla A \cdot \boldsymbol{n} \, dS,
}
and therefore vanishes in a periodic domain or under suitable boundary conditions.

The geometric condition \eqref{intvor} implies the integrability of the vorticity \( \boldsymbol{\omega} \), in the sense of the Frobenius theorem, which locally defines the normal of a two-dimensional surface. We emphasize that any two-dimensional velocity field embedded in a three-dimensional domain satisfies both conditions \eqref{2dNS}, and thus the following Madelung transformation applies to general two-dimensional Navier-Stokes turbulence. 

To see this, note that in 2D we have \( \boldsymbol{\omega} = \omega_z(x, y) \nabla z \), so that \eqref{intvor} follows immediately. On the other hand, the completeness (i.e., the ability to express any 2D flow in the form \eqref{epi2d}) of the Clebsch representation \eqref{epi2d} is shown in \cite{YosClebsch}.

Substituting the representation \eqref{epi2d} into the momentum equation of the Navier-Stokes equations \eqref{NS}, 
we obtain 
\eq{
\nabla&\lrs{\frac{\p A}{\p t}+\frac{p}{\rho}\frac{\p q}{\p t}+\frac{1}{2}m\bol{u}^2-\frac{\hbar^2}{2m}\frac{\Delta\sqrt{\rho}}{\sqrt{\rho}}+V+\eta-m\lr{\frac{4}{3}\nu+\xi}\nabla\cdot\bol{u}}
\\&=\frac{p}{\rho^2}\lrs{\frac{\p\rho}{\p t}+\nabla\cdot\lr{\rho\bol{u}}}\nabla q-\frac{1}{\rho}\lrs{\frac{\p p}{\p t}+\nabla\cdot\lr{p\bol{u}}}\nabla q+\lr{\frac{\p q}{\p t}+\bol{u}\cdot\nabla q}\nabla\lr{\frac{p}{\rho}}-m\nu\nabla\cp\bol{\omega}.\label{NSm}
}
Now suppose that all quantities in the above equation are smooth. 
If $\bol{\omega}=\bol{0}$ in some open region $U\subset\Omega$, then 
$\nabla\cp\bol{\omega}=\bol{0}$ in $U$. 
Hence, the last term in equation \eqref{NSm} can appear only when
$\bol{\omega}=\bol{0}$ in some measure zero subset $\Sigma_0\subset\Omega$, 
or $\bol{\omega}\neq\bol{0}$.  
We may therefore restrict our attention to the open  
region $\tilde{\Omega}=\Omega-\lr{\bar{U}\cup\bar{\Sigma}_0}$ where $\bol{\omega}\neq\bol{0}$,
under the assumption that the value of the fields in $U\cup\Sigma_0$ 
will be eventually determined by regularity and boundary conditions.
The condition $\bol{\omega}\neq\bol{0}$ implies that $\nabla\lr{p/\rho}\cp\nabla q\neq\bol{0}$. Combining this fact with equation \eqref{intvor}, it follows that
\eq{
m\nabla\cp\bol{\omega}=-\alpha\nabla \lr{\frac{p}{\rho}}+\beta\nabla q,
}
for some functions $\alpha\lr{\bol{x},t}$
and $\beta\lr{\bol{x},t}$. 
Solving this equation for $\alpha$ and $\beta$ gives
\sys{
&\alpha=-\frac{\nabla\cp\bol{\omega}\cdot\nabla q\cp\bol{\omega}}{\omega^2}=-\Delta q-\frac{\lrs{\nabla q,\nabla\lr{\frac{p}{\rho}}}\cdot\nabla q\cp\bol{\omega}}{m\omega^2}=-\tilde{\Delta}q,\\
&\beta=-\frac{\nabla\cp\bol{\omega}\cdot\nabla\lr{\frac{p}{\rho}}\cp\bol{\omega}}{\omega^2}=-\Delta\lr{\frac{p}{\rho}}-\frac{\lrs{\nabla q,\nabla\lr{\frac{p}{\rho}}}\cdot\nabla\lr{\frac{p}{\rho}}\cp\bol{\omega}}{m\omega^2}=-\tilde{\Delta}\lr{\frac{p}{\rho}},
}{ab}
where we used standard vector identities, introduced the Lie-bracket $\lrs{\cdot,\cdot}$ acting on vector fields $X,Y$ according to  
\eq{
\lrs{X,Y}=X\cdot\nabla Y-Y\cdot\nabla X,
}
and defined the second-order differential operator
\eq{
\tilde{\Delta}=\Delta-\frac{\lrs{\nabla q,\nabla\lr{\frac{p}{\rho}}}\cdot\bol{\omega}\cp\nabla}{m\omega^2}.
}
It follows that the vector field \eqref{epi2d} is a solution of the Navier-Stokes equations \eqref{NS} whenever the Clebsch potentials $A$, $p$, $\rho$, and $q$ 
satisfy the system of equations
\sys{
&\frac{\p A}{\p t}=-\bol{u}\cdot\nabla A+\frac{1}{2}m\bol{u}^2+\frac{\hbar^2}{2m}\frac{\Delta\sqrt{\rho}}{\sqrt{\rho}}-V-\eta+m\lr{\frac{4}{3}\nu+\xi}\nabla\cdot\bol{u}-\nu\frac{p}{\rho}
%\lrs{\Delta q+\frac{\lrs{\nabla q,\nabla\lr{\frac{p}{\rho}}}\cdot\nabla q\cp\bol{\omega}}{m\omega^2}}
\tilde{\Delta}q,\\
&\frac{\p p}{\p t}=-\nabla\cdot\lr{p\bol{u}}
%+\nu\rho\Delta\lr{\frac{p}{\rho}}
+\nu\rho
%\frac{\lrs{\nabla q,\nabla\lr{\frac{p}{\rho}}}\cdot\nabla\lr{\frac{p}{\rho}}\cp\bol{\omega}}{m\omega^2}
\tilde{\Delta}\lr{\frac{p}{\rho}}
,\\
&\frac{\p q}{\p t}=-\bol{u}\cdot\nabla q+\nu
\tilde{\Delta}q
%\frac{\lrs{\nabla q,\nabla\lr{\frac{p}{\rho}}}\cdot\nabla q\cp\bol{\omega}}{m\omega^2}
,\\
&\frac{\p\rho}{\p t}=-\nabla\cdot\lr{\rho\bol{u}}.
}{NS3}
The Madelung transformation for this system requires two additional steps. 
First, introduce new Madelung potentials $\rho_1$, $\rho_2$, $S_1$, and $S_2$ according to
\begin{equation}
\rho=\rho_1+\rho_2,~~~~p=\rho_1-\rho_2,~~~~A=\lr{S_1+S_2}/2,~~~~q=\lr{S_1-S_2}/2.\label{Mad}
\end{equation}
In these variables, the vector  fields $m\bol{u}$ and $m\bol{\omega}$ can be written as
\sys{
&m\bol{u}=\frac{1}{2}\nabla\lr{S_1+S_2}+\frac{1}{2}\frac{\rho_1-\rho_2}{\rho_1+\rho_2}\nabla\lr{S_1-S_2}=\frac{\rho_1\nabla S_1+\rho_2\nabla S_2}{\rho_1+\rho_2},\\
&m\bol{\omega}=\nabla\lr{\frac{\rho_1}{\rho_1+\rho_2}}\cp\nabla S_1+\nabla\lr{\frac{\rho_2}{\rho_1+\rho_2}}\cp\nabla S_2.
}{uom}
Furthermore, 
system \eqref{NS3} takes the form
\begin{subequations}
\begin{align}
&\frac{\p\rho_1}{\p t}=-\nabla\cdot\lr{\rho_1\bol{u}}+\frac{1}{2}\nu\rho\tilde{\Delta}\lr{\frac{p}{\rho}}
%\lrs{\rho\Delta\lr{\frac{p}{\rho}}+\rho\frac{\lrs{\nabla q,\nabla\lr{\frac{p}{\rho}}}\cdot\nabla\lr{\frac{p}{\rho}}\cp\bol{\omega}}{\omega^2}}
,\\
&\frac{\p\rho_2}{\p t}=-\nabla\cdot\lr{\rho_2\bol{u}}-\frac{1}{2}\nu\rho\tilde{\Delta}\lr{\frac{p}{\rho}}
%\lrs{\rho\Delta\lr{\frac{p}{\rho}}+\rho\frac{\lrs{\nabla q,\nabla\lr{\frac{p}{\rho}}}\cdot\nabla\lr{\frac{p}{\rho}}\cp\bol{\omega}}{\omega^2}}
,\\
&\frac{\p S_1}{\p t}=-\bol{u}\cdot\nabla S_1+\frac{1}{2}m\bol{u}^2-V_q-V-\eta+m\lr{\frac{4}{3}\nu+\xi}\nabla\cdot\bol{u}+\nu\lr{1-\frac{p}{\rho}}\tilde{\Delta}q
%\lrs{\Delta q+\frac{\lrs{\nabla q,\nabla\lr{\frac{p}{\rho}}}\cdot\nabla q\cp\bol{\omega}}{\omega^2}}
,\\
&\frac{\p S_2}{\p t}=-\bol{u}\cdot\nabla S_2+\frac{1}{2}m\bol{u}^2-V_q-V-\eta+m\lr{\frac{4}{3}\nu+\xi}\nabla\cdot\bol{u}-\nu\lr{1+\frac{p}{\rho}}\tilde{\Delta}q
%\lrs{\Delta q-\frac{\lrs{\nabla q,\nabla\lr{\frac{p}{\rho}}}\cdot\nabla q\cp\bol{\omega}}{\omega^2}}
,
\end{align}\label{epi2d_EoM2}
\end{subequations}
where as usual $V_q=-\frac{\hbar^2}{2m}\frac{\Delta\sqrt{\rho}}{\sqrt{\rho}}$. 
The new variables are then used to define a complex valued 2-component spinor field $\bol{\Psi}=\lr{\Psi_1,\Psi_2}$ with 
\begin{equation}
\Psi_i=\sqrt{\rho_i}\exp\lrc{{{\rm i}\frac{S_i}{\hbar}}},~~~~i=1,2.
\end{equation}
The following identities hold for $i=1,2$:
\begin{subequations}
\begin{align}
&\rho_i=\Psi_i\Psi_i^{\ast}=\abs{\Psi_i}^2,\\
&\nabla\rho_i=\Psi_i^{\ast}\nabla\Psi_i+\Psi_i\nabla\Psi_i^{\ast},\\
&\nabla S_i=\frac{{\rm i}\hbar}{2\rho_i}\lr{\Psi_i\nabla\Psi_i^{\ast}-\Psi_i^{\ast}\nabla\Psi_i},\\
&\abs{\nabla S_i}^2=-\frac{\hbar^2}{4}\abs{\nabla\log\rho_i}^2+\frac{\hbar^2}{\rho_i}\nabla\Psi_i\cdot\nabla\Psi_i^{\ast},\\
&m\bol{u}=\frac{{\rm i}\hbar}{2\rho}\lr{\Psi_1\nabla\Psi_1^{\ast}-\Psi_1^{\ast}\nabla\Psi_1+\Psi_2\nabla\Psi_2^{\ast}-\Psi_2^{\ast}\nabla\Psi_2},\\
&\frac{\Delta\sqrt{\rho}}{\sqrt{\rho}}=\frac{\rho_1}{\rho}\frac{\Delta\sqrt{\rho_1}}{\sqrt{\rho_1}}+\frac{\rho_2}{\rho}\frac{\Delta\sqrt{\rho_2}}{\sqrt{\rho_2}}+\frac{1}{4}\lr{\frac{\rho_1}{\rho}\abs{\nabla\log\rho_1}^2+\frac{\rho_2}{\rho}\abs{\nabla\log\rho_2}^2-\abs{\nabla\log\rho}^2}.
\end{align}\label{id2}
\end{subequations}
as well as
\eq{
{\rm i}\hbar\frac{\p\Psi_i}{\p t}=\lr{\frac{{\rm i}\hbar}{2\rho_i}\frac{\p\rho_i}{\p t}-\frac{\p S_i}{\p t}}\Psi_i.\label{82}
}
Using \eqref{id2} and \eqref{82}, one can convert system \eqref{epi2d_EoM2} into the coupled equations
\begin{equation}
\begin{split}
{\rm i}\hbar\frac{\p\Psi_1}{\p t}=&
-\frac{\hbar^2}{2m}\frac{\Delta\sqrt{\rho}}{\sqrt{\rho}}\Psi_1
+\lr{
V+\eta}\Psi_1\\
&+\lrs{-\frac{{\rm i}\hbar}{2\rho_1}\nabla\cdot\lr{\rho_1\bol{u}}+\bol{u}\cdot\nabla S_1-\frac{1}{2}m\bol{u}^2}\Psi_1\\
&+\nu\frac{{\rm i}\hbar}{4}\frac{\rho}{\rho_1}\lrs{\tilde{\Delta}\lr{\frac{p}{\rho}}}\Psi_1-\lrs{m\lr{\frac{4}{3}\nu+\xi}\nabla\cdot\bol{u}+\nu\lr{1-\frac{p}{\rho}}\tilde{\Delta}q}\Psi_1,\label{psi1}
%\lr{\Psi_1\Psi_2\nabla\Psi_1^{\ast}\cdot\nabla\Psi_2^{\ast}-
%\Psi_1\Psi_2^{\ast}\nabla\Psi_1^{\ast}\cdot\nabla\Psi_2
%-\Psi_1^{\ast}\Psi_2\nabla\Psi_1^\cdot\nabla\Psi_2^{\ast}
%+\Psi_1^{\ast}\Psi_2^{\ast}\nabla\Psi_1\cdot\nabla\Psi_2
%}}\Psi_1,
\end{split}
\end{equation}
and 
\begin{equation}
\begin{split}
{\rm i}\hbar\frac{\p\Psi_2}{\p t}=&
-\frac{\hbar^2}{2m}\frac{\Delta\sqrt{\rho}}{\sqrt{\rho}}\Psi_2
+
\lr{
V+\eta}\Psi_2\\
&+\lrs{-\frac{{\rm i}\hbar}{2\rho_2}\nabla\cdot\lr{\rho_2\bol{u}}+\bol{u}\cdot\nabla S_2-\frac{1}{2}m\bol{u}^2}\Psi_2\\
&-\nu\frac{{\rm i}\hbar}{4}\frac{\rho}{\rho_2}\lrs{\tilde{\Delta}\lr{\frac{p}{\rho}}}\Psi_2-\lrs{m\lr{\frac{4}{3}\nu+\xi}\nabla\cdot\bol{u}-\nu\lr{1+\frac{p}{\rho}}\tilde{\Delta}q}\Psi_2.\label{psi2}
%\lr{\Psi_1\Psi_2\nabla\Psi_1^{\ast}\cdot\nabla\Psi_2^{\ast}-
%\Psi_1\Psi_2^{\ast}\nabla\Psi_1^{\ast}\cdot\nabla\Psi_2
%-\Psi_1^{\ast}\Psi_2\nabla\Psi_1^\cdot\nabla\Psi_2^{\ast}
%+\Psi_1^{\ast}\Psi_2^{\ast}\nabla\Psi_1\cdot\nabla\Psi_2
%}}\Psi_1,
\end{split}
\end{equation}
Note that in equations \eqref{psi1} and \eqref{psi2}, we have not expressed all terms with the wave functions \( \Psi_1 \) and \( \Psi_2 \) to simplify the notation. However, these explicit expressions can be readily derived using \eqref{Mad} and \eqref{id2}.

Equations \eqref{psi1} and \eqref{psi2} represent the `quantum' form of the Navier-Stokes flows \eqref{NS} governed by \eqref{2dNS}. It is important to emphasize again that these equations are applicable to general 2D flows.

The Madelung transformation for flows obeying \eqref{2dNS} can be generalized to fully 3D flows. In this case, the velocity field \( \boldsymbol{u} \) has the Clebsch form
\eq{
\boldsymbol{u} = \frac{1}{m}\nabla A + \frac{p}{m\rho} \nabla q + \frac{\tau}{m\rho} \nabla \sigma,
}
where $\tau\lr{\bol{x},t}$ and $\sigma\lr{\bol{x},t}$ denote an additional pair of Clebsch potentials. 
In this setting, the treatment of the viscous terms on the right-hand side of the Navier-Stokes equations requires a distinction between regions where \( \boldsymbol{\omega} \cdot \nabla \times \boldsymbol{\omega} = 0 \) and regions where \( \boldsymbol{\omega} \cdot \nabla \times \boldsymbol{\omega} \neq 0 \) (on this point, see \cite{Sato21}). The quantum form of the Navier-Stokes equations can then be derived by introducing the Madelung potentials \( \rho_1, \rho_2, \rho_3, S_1, S_2, S_3 \). This construction follows the same logic as the transformation for flows obeying \eqref{2dNS}, though it involves more intricate algebraic steps.

\section{Concluding Remarks}

In this paper, we have explored the intriguing connection between quantum mechanics and fluid dynamics by investigating whether a fluid system with internal rotational (spin) degrees of freedom can be described by the Schrödinger equation. Our approach was grounded in deriving the governing equations from fundamental fluid mechanical principles, without presupposing quantum corrections.

In the nonrelativistic case, we demonstrated that incorporating spin introduces a quantum correction to the classical fluid energy, leading naturally to the Schrödinger equation in Madelung form when coupled with Maxwell's equations. Extending this formalism to relativistic systems, we identified the stress-energy-momentum tensor and showed that the quantum correction is associated with the energy density of an oscillator, its frequency determined by the vorticity of spin motion. This result allowed us to establish a link between the Ricci scalar curvature, governed by Einstein’s field equations, and the fluid density in a static, spherically symmetric configuration.

Furthermore, we generalized the Madelung transformation to compressible Navier-Stokes flows with vorticity and viscosity, employing a tailored Clebsch representation of the velocity field. This formulation facilitates the development of wave functions that encode the dynamics of such flows, offering a promising avenue for studying more complex fluid systems.

The quantum-fluid correspondence framework presented here has the potential to deepen our understanding of fluid systems with internal spin, especially in astrophysical contexts where such models are often applicable. Additionally, this work offers insights into the potential coupling of gravity and spin in fluid-like systems, with possible applications in both classical and quantum mechanics.

Finally, while our analysis provides a foundation for further exploration of quantum behavior in fluid systems, we also acknowledge the challenges in extending this framework to quantum gravity. The inherent nonlinearity introduced by the interaction between the wave function and the spacetime metric contrasts with the linear nature of quantum mechanics. Moreover, issues related to the collapse of the wave function and its effects on spacetime geometry pose significant theoretical challenges that remain to be addressed.

%\appendix

\section*{Statements and Declarations}
\subsection*{Data availability}
Data sharing not applicable to this article as no datasets were generated or analysed during the current study.

\subsection*{Funding}
The research of NS was partially supported by JSPS KAKENHI Grant No.  22H04936 and No. 24K00615. 
%This work was partly supported by MEXT Promotion of Distinctive Joint Research Center Program JPMXP0723833165. 

\subsection*{Competing interests} 
The authors have no competing interests to declare that are relevant to the content of this article.

\end{document}